\documentclass[aps,10pt,prl,twocolumn,showpacs,amsmath,amssymb,superscriptaddress,longbibliography]{revtex4-1}

\usepackage{graphicx,bm,mathptmx} 
\usepackage[dvipsnames]{xcolor}
\usepackage{colordvi}

\usepackage{hyperref}

\hypersetup{
    colorlinks,
    breaklinks=true,
    linkcolor={red!50!black},
    citecolor={blue!50!black},
    urlcolor={blue!80!black},
    pdfpagemode=UseNone, 
    bookmarksopen=false,
    pdfstartview=FitH,
    pdfview=FitH
}

\def\ForrestGreen#1{\textcolor{forrestgreen}{#1}}
\def\Blue#1{\textcolor{blue}{#1}}

\def\ket#1{|#1\rangle}
\def\C{\mathbb{C}}
\def\F{\mathbb{F}}

\definecolor{orange}{rgb}{1,0.5,0}
\definecolor{forrestgreen}{cmyk}{0.91,0,0.88,0.12}

\begin{document}

\title{Entanglement-Assisted Quantum Communication Beating the Quantum
Singleton Bound}

\author{Markus~Grassl}
\affiliation{International Centre for Theory of Quantum Technologies,
  University Gdansk, 80-308 Gdansk, Poland}
\affiliation{Max-Planck-Institut f\"ur die Physik des Lichts, 
91058 Erlangen, Germany} 

\begin{abstract}
Brun, Devetak, and Hsieh [Science \textbf{314}, 436 (2006)]
demonstrated that pre-shared entanglement between sender and receiver
enables quantum communication protocols that have better parameters
than schemes without the assistance of entanglement.  Subsequently,
the same authors derived a version of the so-called quantum Singleton
bound that relates the parameters of the entanglement-assisted
quantum-error correcting codes proposed by them.  We present a new
entanglement-assisted quantum communication scheme with parameters
violating this bound in certain ranges.
\end{abstract}


\maketitle

\emph{Introduction.}---Entanglement is a resource that enables or
enhances many tasks in quantum communication. When sender and receiver
share a maximally entangled state, quantum teleportation allows the
sender to transmit an unknown quantum state by just sending a finite
amount of classical information over a noiseless classical channel
\cite{Teleportation}.  When the entangled states are initially
distributed over a noisy quantum channel, using local quantum
operations and a noiseless classical channel, sender and receiver can
extract a smaller number of maximally entangled states with higher
fidelity by a distillation process \cite{BBPS96}.  The correspondence
between entanglement distillation protocols and quantum
error-correcting codes in a communication scenario has been discussed
in \cite{BDSW96}. Quantum error-correcting codes (QECCs), however, are
somewhat more general in the sense that they allow to recover a
quantum state affected by a general quantum channel, provided that a
suitable encoding for that channel exists \cite{KnLa97}.  Hence QECCs
can be used both for communication and storage, and they are essential
ingredients for fault-tolerant quantum computation \cite{ShorFT96}.
On the other hand, in \cite{BDH06} it has been shown that the
performance of QECCs in a communication scenario can be improved when
a noisy quantum channel is assisted by entanglement.

In this Letter we present a quantum communication scheme that also
uses a noisy quantum channel assisted by entanglement.  The main idea
it to execute a teleportation protocol in which the classical
information is protected using a code and then sent via the noisy
quantum channel to the receiver.  This allows to use classical
error-correcting codes.  In some range, our scheme has better
parameters than the one proposed in \cite{BDH06}, showing that
the adaption of the quantum Singleton bound to that class of codes
presented in \cite{BDH14} can be violated.

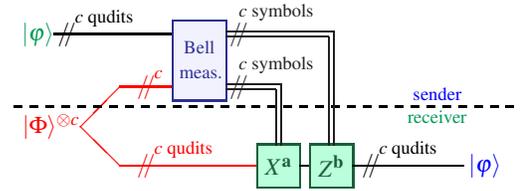
\begin{figure}[hbt]
\centerline{{\begin{picture}(190,70)(10,10)
\put(35,35){\textcolor{red}{\makebox(0,0)[r]{$\ket{\Phi}^{\otimes c}$}}}
\put(35,35){\textcolor{red}{\line(1,1){15}}}
\put(35,35){\textcolor{red}{\line(1,-1){15}}}
\put(60,50){\textcolor{red}{\makebox(0,0){$/\!\!/$}}}
\put(63,53){\textcolor{red}{\makebox(0,0)[lb]{\scriptsize$c$}}}
\put(60,20){\textcolor{red}{\makebox(0,0){$/\!\!/$}}}
\put(63,23){\textcolor{red}{\makebox(0,0)[lb]{\scriptsize$c$ qudits}}}
\put(50,50){\textcolor{red}{\line(1,0){20}}}
\put(50,20){\textcolor{red}{\line(1,0){52}}}
\put(102,12){\color[rgb]{0.73,1,0.86}{\rule{16\unitlength}{16\unitlength}}}
\put(102,12){\color[rgb]{0.1,0.37,0.23}{\framebox(16,16){$X^{\mathbf{a}}$}}}
\put(118,20){\line(1,0){4}}
\put(122,12){\color[rgb]{0.73,1,0.86}{\rule{16\unitlength}{16\unitlength}}}
\put(122,12){\color[rgb]{0.1,0.37,0.23}{\framebox(16,16){$Z^{\mathbf{b}}$}}}
\put(145,20){\makebox(0,0){$/\!\!/$}}
\put(148,23){\makebox(0,0)[lb]{\scriptsize$c$ qudits}}
\put(138,20){\line(1,0){42}}
\put(182,20){\makebox(0,0)[l]{\Blue{$\ket{\varphi}$}}}
\put(25,70){\makebox(0,0)[r]{\ForrestGreen{$\ket{\varphi}$}}}
\put(30,70){\makebox(0,0){$/\!\!/$}}
\put(33,73){\makebox(0,0)[lb]{\scriptsize$c$ qudits}}
\put(25,70){\line(1,0){45}}
\multiput(10,42.5)(6,0){32}{\line(1,0){3}}
\put(170,45){\makebox(0,0)[b]{\scriptsize\Blue{sender}}}
\put(170,41){\makebox(0,0)[t]{\scriptsize\ForrestGreen{receiver}}}
\put(0,0){\textBlue}
\put(70,45){\color[rgb]{0.95,0.95,1}{\rule{20\unitlength}{30\unitlength}}}
\put(70,45){\framebox(20,30){}}
\put(80,65){\makebox(0,0)[c]{\scriptsize Bell}}
\put(80,55){\makebox(0,0)[c]{\scriptsize meas.\!}}
\put(0,0){\textBlack}
\put(95,70){\makebox(0,0){$/\!\!/$}}
\put(95,75){\makebox(0,0)[lb]{\scriptsize$c$ symbols}}
\put(90,71){\line(1,0){41}}
\put(90,69){\line(1,0){39}}
\put(131,71){\line(0,-1){43}}
\put(129,69){\line(0,-1){41}}
\put(95,50){\makebox(0,0){$/\!\!/$}}
\put(95,55){\makebox(0,0)[lb]{\scriptsize$c$ symbols}}
\put(90,51){\line(1,0){21}}
\put(90,49){\line(1,0){19}}
\put(111,51){\line(0,-1){23}}
\put(109,49){\line(0,-1){21}}
\end{picture}}}

\caption{Teleportation protocol. The maximally entangled states
  $\ket{\Phi}^{\otimes c}$ can be prepared by either party, or even a
  third party.}
\label{fig:teleportation}
\end{figure}

\emph{Quantum Teleportation.}---We start with a short summary quantum
teleportation \cite{Teleportation}, as illustrated in
Fig.~\ref{fig:teleportation}.

The aim is to transmit an arbitrary quantum state $\ket{\varphi}$ in
the Hilbert space $\mathcal{H}\cong(\C^d)^{\otimes c}$ of $c$ quantum
systems of dimension $d$ (\emph{qudits}).  The protocol is assisted by
$c$ copies of a maximally entangled bipartite state
\begin{equation}\label{eq:maxentanglement}
\ket{\Phi}_{SR}=\frac{1}{\sqrt{d}}\sum_{i=0}^{d-1}\ket{i}_S\ket{i}_R
\end{equation}
which are shared by sender $S$ and receiver $R$.  Applying
Heisenberg-Weyl operators $X^aZ^b$ (or generalized Pauli matrices) on
one of the systems, the collection of the resulting states constitutes
the generalized Bell basis
\begin{equation}\label{eq:BellBasis}
\{\ket{\Phi^{a,b}}=(I\otimes X^aZ^b)\ket{\Phi}\colon a,b=0,\ldots d-1\}.
\end{equation}
Here $X$ is a cyclic shift operator in the standard basis and $Z$ its
diagonal form.

The sender performs a generalized Bell measurement, i.\,e., a
measurement in the generalized Bell basis \eqref{eq:BellBasis}, on the
input state $\ket{\varphi}$ and $c$ qudits from the maximally
entangled states. For each pair of qudits, one obtains a pair
$(a_i,b_i)$ of classical values.  The strings
$\mathbf{a}=(a_1,a_2,\ldots,a_c)$ and
$\mathbf{b}=(b_1,b_2,\ldots,b_c)$ with $c$ classical symbols each are
sent to the receiver, indicated by the double lines in
Fig.~\ref{fig:teleportation}. Depending on the values of $\mathbf{a}$
and $\mathbf{b}$, the receiver applies correction operations
$X^{\mathbf{a}}$ and $Z^{\mathbf{b}}$ and obtains the original state
$\ket{\varphi}$.

\emph{Quantum Error-Correcting Codes.}---A standard quantum
error-correcting code $\mathcal{C}$ of length $n$ is a subspace of the
Hilbert space $(\C^d)^{\otimes n}$ of $n$ qudits.  Usually, $d$ is
assumed to be a power of a prime, i.e., $d=q=p^m$ for some prime $p$.
A QECC encoding $k$ qudits has dimension $q^k$ and is denoted by
$[\![n,k,d]\!]_q$.  A QECC with minimum distance $d=2t+1$ allows to
correct all errors affecting no more than $t$ of the subsystems.  When
the position of the errors is known, then errors on up to $d-1$
subsystems can be corrected \cite{GBP97}.  Alternatively, all errors
on no more than $d-1$ of the subsystems that act non-trivially on the
code can be detected.  Independent of the dimension $q$ of the
subsystems, the parameters of a QECC are constraint by the so-called
quantum Singleton bound \cite{KnLa97,Rai99}
\begin{equation}\label{eq:QuantumSingletonBound}
2d \le n-k+2.
\end{equation}
Codes meeting this bound with equality are called \emph{quantum MDS
  (QMDS) codes}.  With few exceptions, QMDS codes are only known for
length $n\le q^2+1$ and minimum distance $d\le q+1$ \cite{GrRo15}.

\begin{figure}[hbt]
\centerline{{\begin{picture}(235,65)(2.5,40)
\put(25,80){\makebox(0,0)[r]{$\ket{0^{n-k}}$}}
\put(30,80){\makebox(0,0){$/\!\!/$}}
\put(31,85){\makebox(0,0)[lb]{\scriptsize$n-k$ qudits}}
\put(25,80){\line(1,0){45}}
\put(25,60){\makebox(0,0)[r]{\Blue{$\ket{\varphi}$}}}
\put(30,60){\makebox(0,0){$/\!\!/$}}
\put(33,63){\makebox(0,0)[lb]{\scriptsize$k$ qudits}}
\put(25,60){\line(1,0){45}}
\put(0,0){\textBlue}
\put(70,55){\color[rgb]{0.95,0.95,1}{\rule{20\unitlength}{30\unitlength}}}
\put(70,55){\framebox(20,30){$\mathcal{E}$}}
\put(0,0){\textBlack}
\put(95,70){\makebox(0,0){$/\!\!/$}}
\put(95,75){\makebox(0,0)[lb]{\scriptsize$n$ qudits}}
\put(90,70){\line(1,0){35}}
\put(130,98){\makebox(0,0)[r]{\scriptsize\Blue{sender}}}
\multiput(135,83)(0,6){4}{\line(0,1){3}}
\put(140,98){\makebox(0,0)[l]{\scriptsize\ForrestGreen{receiver}}}
\put(0,0){\textMagenta}
\put(125,60){\color[rgb]{1,0.95,1}{\rule{20\unitlength}{20\unitlength}}}
\put(125,60){\framebox(20,20){$\mathcal{N}$}}
\put(0,0){\textBlack}
\multiput(135,57)(0,-6){3}{\line(0,-1){3}}
\put(145,70){\line(1,0){20}}
\put(155,70){\makebox(0,0){$/\!\!/$}}
\put(158,73){\makebox(0,0)[lb]{\scriptsize$n$}}
\put(165,60){\color[rgb]{0.73,1,0.86}{\rule{20\unitlength}{20\unitlength}}}
\put(165,60){\color[rgb]{0.1,0.37,0.23}{\framebox(20,20){$\mathcal{D}$}}}
\put(190,70){\makebox(0,0){$/\!\!/$}}
\put(193,73){\makebox(0,0)[lb]{\scriptsize$k$ qudits}}
\put(185,70){\line(1,0){35}}
\put(222,70){\makebox(0,0)[l]{\ForrestGreen{$\ket{\varphi'}$}}}
\end{picture}}}

\caption{Scheme of a communication protocol using a quantum
  error-correcting code $[\![n,k,d]\!]_q$.} 
\label{fig:QECC}
\end{figure}
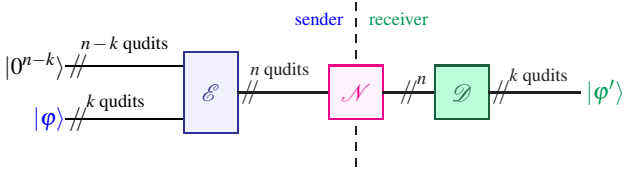

The overall scheme is illustrated in Fig.~\ref{fig:QECC}. The unitary
encoding operator $\mathcal{E}$ maps the input state $\ket{\varphi}$
with $k$ qudits and $n-k$ ancilla qudits $\ket{0^{n-k}}$ to the encoding
space with $n$ qudits. Those $n$ qudits are sent over a noisy quantum
channel $\mathcal{N}$, whose output enters the decoder $\mathcal{D}$.
The decoder outputs a quantum state $\ket{\varphi'}$ with $k$ qudits.
When the error can be corrected, the states $\ket{\varphi'}$ and
$\ket{\varphi}$ are equal.

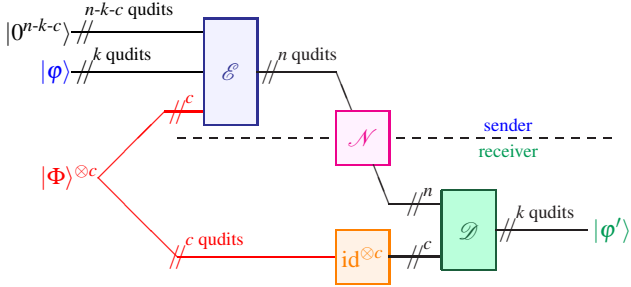
\begin{figure}[hbt]
\centerline{{\begin{picture}(245,103)(-5,0)
\put(35,40){\textcolor{red}{\makebox(0,0)[r]{$\ket{\Phi}^{\otimes c}$}}}
\put(35,40){\textcolor{red}{\line(1,1){25}}}
\put(35,40){\textcolor{red}{\line(1,-1){30}}}
\put(60,65){\textcolor{red}{\line(1,0){15}}}
\put(65,65){\textcolor{red}{\makebox(0,0){$/\!\!/$}}}
\put(68,68){\textcolor{red}{\makebox(0,0)[lb]{\scriptsize$c$}}}
\put(65,10){\textcolor{red}{\line(1,0){60}}}
\put(65,10){\textcolor{red}{\makebox(0,0){$/\!\!/$}}}
\put(68,13){\textcolor{red}{\makebox(0,0)[lb]{\scriptsize$c$ qudits}}}
\put(25,95){\makebox(0,0)[r]{$\ket{0^{n\text{-}k\text{-}c}}$}}
\put(30,95){\makebox(0,0){$/\!\!/$}}
\put(30,100){\makebox(0,0)[lb]{\scriptsize$n\text{-}k\text{-}c$ qudits}}
\put(25,95){\line(1,0){50}}
\put(25,80){\makebox(0,0)[r]{\Blue{$\ket{\varphi}$}}}
\put(30,80){\makebox(0,0){$/\!\!/$}}
\put(33,83){\makebox(0,0)[lb]{\scriptsize$k$ qudits}}
\put(25,80){\line(1,0){50}}
\put(0,0){\textBlue}
\put(75,60){\color[rgb]{0.95,0.95,1}{\rule{20\unitlength}{40\unitlength}}}
\put(75,60){\framebox(20,40){$\mathcal{E}$}}
\put(0,0){\textBlack}
\put(100,80){\makebox(0,0){$/\!\!/$}}
\put(103,83){\makebox(0,0)[lb]{\scriptsize$n$ qudits}}
\put(95,80){\line(1,0){30}}
\put(125,80){\line(2,-5){6}}
\put(0,0){\textMagenta}
\put(125,45){\color[rgb]{1,0.95,1}{\rule{20\unitlength}{20\unitlength}}}
\put(125,45){\framebox(20,20){$\mathcal{N}$}}
\put(0,0){\textBlack}
\put(145,30){\line(-2,5){6}}
\multiput(148,55)(6,0){14}{\line(1,0){3}}
\multiput(122,55)(-6,0){10}{\line(-1,0){3}}
\put(190,57.5){\makebox(0,0)[b]{\scriptsize\Blue{sender}}}
\put(190,52.5){\makebox(0,0)[t]{\scriptsize\ForrestGreen{receiver}}}
\put(155,30){\makebox(0,0){$/\!\!/$}}
\put(158,33){\makebox(0,0)[lb]{\scriptsize$n$}}
\put(145,30){\line(1,0){20}}
\put(125,00){\color[rgb]{1,0.95,0.9}{\rule{20\unitlength}{20\unitlength}}}
\put(125,00){\textcolor{orange}{\framebox(20,20){$\text{id}^{\otimes c}$}}}
\put(155,10){\makebox(0,0){$/\!\!/$}}
\put(158,13){\makebox(0,0)[lb]{\scriptsize$c$}}
\put(145,10){\line(1,0){20}}
\put(165,05){\color[rgb]{0.73,1,0.86}{\rule{20\unitlength}{30\unitlength}}}
\put(165,05){\color[rgb]{0.1,0.37,0.23}{\framebox(20,30){$\mathcal{D}$}}}
\put(190,20){\makebox(0,0){$/\!\!/$}}
\put(193,23){\makebox(0,0)[lb]{\scriptsize$k$ qudits}}
\put(185,20){\line(1,0){35}}
\put(222,20){\makebox(0,0)[l]{\ForrestGreen{$\ket{\varphi'}$}}}
\end{picture}}}

\caption{Communication scheme using an entanglement-assisted quantum
  code $[\![n,k,d;c]\!]_q$.}
\label{fig:EAQECC}
\end{figure}

\emph{Entanglement Assisted Quantum Error-Correcting Codes.}--- An
entanglement assisted quantum error-correcting code (EAQECC), denoted
by $[\![n,k,d;c]\!]_q$, is a quantum error-correcting code that
additionally uses $c$ maximally entangled states. The scheme is
illustrated in Fig.~\ref{fig:EAQECC}.

One half of each maximally entangled state
$\ket{\Phi}\in\C^q\otimes\C^q$ enters the encoding operation
$\mathcal{E}$ together with the $k$-qudit state $\ket{\varphi}$ to be
transmitted and $n-k-c$ ancilla qudits.  The other half of each
maximally entangled state is assumed to be transmitted error-free.
The $n$ qudits output by the encoding operation are sent through the
noisy quantum channel $\mathcal{N}$.  The receiver applies the
decoding operation $\mathcal{D}$ to both the $n$ qudits output from
the channel and the $c$ noiseless qudits from the maximally entangled
states.  

In \cite{BDH14}, the authors formulate a Singleton bound for the
parameters $[\![n,k,d;c]\!]_q$ of an EAQECC:
\begin{equation}\label{eq:EASingletonBound}
2d \le n-k+2+c.
\end{equation}
This bound can be derived by considering the $n$ qudits sent over the
noisy quantum channel $\mathcal{N}$ together with the $c$ qubits from
the maximally entangled states sent over a noiseless channel as a
standard QECC of length $n+c$ in the quantum Singleton bound
\eqref{eq:QuantumSingletonBound}. This, however, ignores the fact that
the $c$ additional qudits are assumed to be error-free. The approach
in \cite{LaAs18} accounts for this additional assumption. In Theorem~6
of \cite{LaAs18}, the bound \eqref{eq:EASingletonBound} has been
shown to be valid for the case $d\le (n+2)/2$.

In \cite{FCX16}, EAQECCs meeting the bound (\ref{eq:EASingletonBound})
with equality were constructed. Some of the codes use only $c=1$ or
$c=2$ maximally entangled states, the maximal value is $c=q+1$.  While
the length of the codes is bounded by $q^2+1$ like in the case of QMDS
codes, the minimum distance can be as large as $d=2q$ for some of the
codes, compared to $d\le q+1$ for most QMDS codes without entanglement
assistance.

In \cite{BDH06}, a construction of EAQECC from any linear code
$[n,\kappa,d]_{q^2}$ over the finite field $\F_{q^2}$ with $q^2$ was
given. The parameters of the resulting EAQECC are
$[\![n,2\kappa-n+c,d;c]\!]_q$, where the number $c$ of maximally
entangled states depends on the classical code and is at most
$n-\kappa$. Using a result from \cite{CMTQR18} it follows that for
$q\ge 3$, any classical linear code $[n,\kappa,d]_{q^2}$ can be
converted into an EAQECC that requires the maximal amount of
entanglement $c=n-\kappa$. The EAQECC has parameters
$[\![n,\kappa,d;n-\kappa]\!]_q$.  From a classical MDS code
$[n,\kappa,n-\kappa+1]_{q^2}$, we obtain an EAQECC with parameters
$[\![n,k,n-k+1;n-k]\!]_q$, meeting the bound
\eqref{eq:EASingletonBound} with equality.  Assuming the maximal value
for $c=n-k$, the minimum distance of an EAQECC from this construction
obeys the bound
\begin{equation}\label{eq:absoluteBound}
d \le n-k+1.
\end{equation}
which is exactly the Singleton bound for classical codes. 

The bound
\eqref{eq:absoluteBound} is also an absolute bound on the minimum
distance of any quantum code for the following reason.  A quantum code
with minimum distance $d$ can correct errors that affect $d-1$ errors
at known positions. Hence, after tracing out $d-1$ of the systems, we
will still be able to recover any encoded state of $k$ qudits. The
residual state has $n-d+1$ qudits, and hence the bound $k\le n-d+1$
follows. 

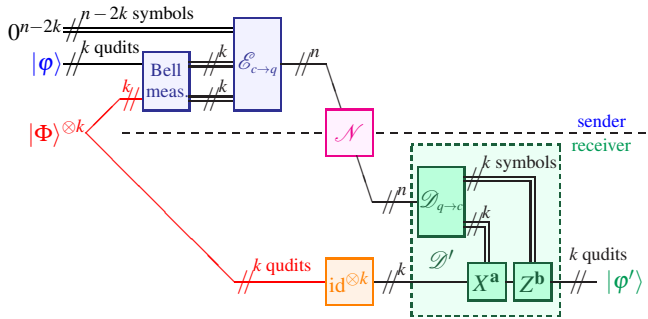
\begin{figure}[hbt]
\unitlength0.8666\unitlength
\centerline{\begin{picture}(280,135)(0,5)
\put(35,85){\textcolor{red}{\makebox(0,0)[r]{$\ket{\Phi}^{\otimes k}$}}}
\put(35,85){\textcolor{red}{\line(1,1){15}}}
\put(35,85){\textcolor{red}{\line(1,-1){65}}}
\put(50,100){\textcolor{red}{\line(1,0){10}}}
\put(55,100){\textcolor{red}{\makebox(0,0){$/\!\!/$}}}
\put(51,103){\textcolor{red}{\makebox(0,0)[lb]{\scriptsize$k$}}}
\put(100,20){\textcolor{red}{\line(1,0){40}}}
\put(105,20){\textcolor{red}{\makebox(0,0){$/\!\!/$}}}
\put(108,23){\textcolor{red}{\makebox(0,0)[lb]{\scriptsize$k$ qudits}}}
\put(25,115){\makebox(0,0)[r]{\Blue{$\ket{\varphi}$}}}
\put(30,115){\makebox(0,0){$/\!\!/$}}
\put(33,118){\makebox(0,0)[lb]{\scriptsize$k$ qudits}}
\put(25,115){\line(1,0){35}}
\put(25,130){\makebox(0,0)[r]{$0^{n-2k}$}}
\put(30,130){\makebox(0,0){$/\!\!/$}}
\put(33,133){\makebox(0,0)[lb]{\scriptsize$n-2k$ symbols}}
\put(25,131){\line(1,0){75}}
\put(25,129){\line(1,0){75}}
\put(0,0){\textBlue}
\put(60,95){\color[rgb]{0.95,0.95,1}{\rule{20\unitlength}{25\unitlength}}}
\put(60,95){\framebox(20,25){}}
\put(70,112.5){\makebox(0,0)[c]{\scriptsize Bell}}
\put(70,102.5){\makebox(0,0)[c]{\scriptsize meas.\!}}
\put(0,0){\textBlack}
\put(90,115){\makebox(0,0){$/\!\!/$}}
\put(93,118){\makebox(0,0)[lb]{\scriptsize$k$}}
\put(80,116){\line(1,0){20}}
\put(80,114){\line(1,0){20}}
\put(90,100){\makebox(0,0){$/\!\!/$}}
\put(93,103){\makebox(0,0)[lb]{\scriptsize$k$}}
\put(80,101){\line(1,0){20}}
\put(80,99){\line(1,0){20}}
\put(0,0){\textBlue}
\put(100,95){\color[rgb]{0.95,0.95,1}\rule{20\unitlength}{40\unitlength}}
\put(100,95){\framebox(20,40){$\mathcal{E}_{\text{\tiny$\scriptstyle c\to q$}}$}}
\put(0,0){\textBlack}
\put(120,115){\line(1,0){20}}
\put(130,115){\makebox(0,0){$/\!\!/$}}
\put(133,118){\makebox(0,0)[lb]{\scriptsize$n$}}
\put(120,115){\line(1,0){20}}
\put(140,115){\line(1,-3){6.666666}}
\put(0,0){\textMagenta}
\put(140,75){\color[rgb]{1,0.95,1}{\rule{20\unitlength}{20\unitlength}}}
\put(140,75){\framebox(20,20){{$\mathcal{N}$}}}
\put(0,0){\textBlack}
\put(160,55){\line(-1,3){6.666666}}
\put(177,5){\color[rgb]{0.92,1,0.96}{\rule{65\unitlength}{75\unitlength}}}
\put(177,5){\color[rgb]{0.1,0.37,0.23}{\dashbox{2}(65,75){}}}
\put(190,30){\color[rgb]{0.1,0.37,0.23}{\makebox(0,0){$\mathcal{D}'$}}}
\multiput(162,85)(6,0){20}{\line(1,0){3}}
\multiput(138,85)(-6,0){15}{\line(-1,0){3}}
\put(260,87.5){\makebox(0,0)[b]{\scriptsize\Blue{sender}}}
\put(260,82.5){\makebox(0,0)[t]{\scriptsize\ForrestGreen{receiver}}}
\put(168,55){\makebox(0,0){$/\!\!/$}}
\put(171,58){\makebox(0,0)[lb]{\scriptsize$n$}}
\put(160,55){\line(1,0){20}}
\put(160,20){\line(1,0){42}}
\put(140,10){\color[rgb]{1,0.95,0.9}{\rule{20\unitlength}{20\unitlength}}}
\put(140,10){\textcolor{orange}{\framebox(20,20){$\text{\footnotesize id}^{\otimes k}$}}}
\put(168,20){\makebox(0,0){$/\!\!/$}}
\put(171,23){\makebox(0,0)[lb]{\scriptsize$k$}}
\put(180,40){\color[rgb]{0.73,1,0.86}{\rule{20\unitlength}{30\unitlength}}}
\put(180,40){\color[rgb]{0.1,0.37,0.23}{\framebox(20,30){$\mathcal{D}_{\text{\tiny$\scriptstyle q\to c$}}$}}}
\put(205,45){\makebox(0,0){$/\!\!/$}}
\put(208,48){\makebox(0,0)[lb]{\scriptsize$k$}}
\put(200,46){\line(1,0){11}}
\put(200,44){\line(1,0){9}}
\put(211,46){\line(0,-1){18}}
\put(209,44){\line(0,-1){16}}
\put(205,65){\makebox(0,0){$/\!\!/$}}
\put(208,68){\makebox(0,0)[lb]{\scriptsize$k$ symbols}}
\put(200,66){\line(1,0){31}}
\put(200,64){\line(1,0){29}}
\put(231,66){\line(0,-1){38}}
\put(229,64){\line(0,-1){36}}
\put(202,12){\color[rgb]{0.73,1,0.86}{\rule{16\unitlength}{16\unitlength}}}
\put(202,12){\color[rgb]{0.1,0.37,0.23}{\framebox(16,16){$X^{\mathbf{a}}$}}}
\put(218,20){\line(1,0){4}}
\put(222,12){\color[rgb]{0.73,1,0.86}{\rule{16\unitlength}{16\unitlength}}}
\put(222,12){\color[rgb]{0.1,0.37,0.23}{\framebox(16,16){$Z^{\mathbf{b}}$}}}
\put(250,20){\makebox(0,0){$/\!\!/$}}
\put(246,28){\makebox(0,0)[lb]{\scriptsize$k$ qudits}}
\put(238,20){\line(1,0){20}}
\put(262,20){\makebox(0,0)[l]{\ForrestGreen{\ForrestGreen{$\ket{\varphi'}$}}}}
\end{picture}}
\caption{Our teleportation-based scheme using $c=k$ maximally entangled states.}
\label{fig:OurScheme1}
\end{figure}

\emph{The New Scheme.}---In our scheme, we use the $c$ maximally entangled states in a
teleportation protocol to transmit $k=c$ qudits. Each generalized Bell
measurement in the teleportation protocol has $q^2$ possible outcomes,
i.e., we have to send a classical string with $2k$ symbols from an
alphabet of size $q$ to the receiver.  As we allow for $n$ uses of a
quantum channel, we can use a classical code $C$ over an alphabet of
size $q$ encoding $2k$ symbols into $n$ symbols, denoted by
$[n,2k,d]_q$, were again $d$ denotes the minimum distance of the code
(for more details about classical error-correcting codes, see for
example \cite{MS77}). The classical string of length $n$ is mapped to
one of the $q^n$ basis states of the Hilbert space of $n$ qudits and
then sent via the noisy quantum channel $\mathcal{N}$ to the receiver.
The receiver measures the output of the quantum channel in the
computational basis and obtains a classical string of length $n$.
Applying error correction for the classical code $C$, the $2k$ symbols
corresponding to the measurement result from the teleportation protocol
are retrieved.  The measurement and the classical decoder are depicted
together as a quantum-to-classical map $\mathcal{D}_{q\to c}$ in
Fig.~\ref{fig:OurScheme1}.  The receiver applies the corresponding
correction operators $X^{\mathbf{a}}$ and $Z^{\mathbf{b}}$ to the $c$
qudits from the $c$ maximally entangled states and completes the
teleportation protocol.  

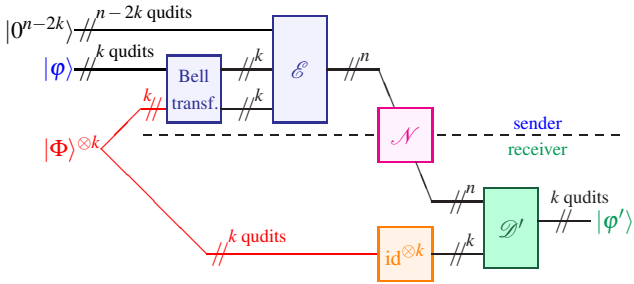
\begin{figure}[hbt]
\centerline{{\begin{picture}(245,110)(-5,5)
\put(35,60){\textcolor{red}{\makebox(0,0)[r]{$\ket{\Phi}^{\otimes k}$}}}
\put(35,60){\textcolor{red}{\line(1,1){15}}}
\put(35,60){\textcolor{red}{\line(1,-1){40}}}
\put(50,75){\textcolor{red}{\line(1,0){10}}}
\put(55,75){\textcolor{red}{\makebox(0,0){$/\!\!/$}}}
\put(51,78){\textcolor{red}{\makebox(0,0)[lb]{\scriptsize$k$}}}
\put(75,20){\textcolor{red}{\line(1,0){65}}}
\put(80,20){\textcolor{red}{\makebox(0,0){$/\!\!/$}}}
\put(83,23){\textcolor{red}{\makebox(0,0)[lb]{\scriptsize$k$ qudits}}}
\put(25,90){\makebox(0,0)[r]{\Blue{$\ket{\varphi}$}}}
\put(30,90){\makebox(0,0){$/\!\!/$}}
\put(33,93){\makebox(0,0)[lb]{\scriptsize$k$ qudits}}
\put(25,90){\line(1,0){35}}
\put(25,105){\makebox(0,0)[r]{$\ket{0^{n-2k}}$}}
\put(30,105){\makebox(0,0){$/\!\!/$}}
\put(33,108){\makebox(0,0)[lb]{\scriptsize$n-2k$ qudits}}
\put(25,105){\line(1,0){75}}
\put(0,0){\textBlue}
\put(60,70){\color[rgb]{0.95,0.95,1}{\rule{20\unitlength}{25\unitlength}}}
\put(60,70){\framebox(20,25){}}
\put(70,87.5){\makebox(0,0)[c]{\scriptsize Bell}}
\put(70,77.5){\makebox(0,0)[c]{\scriptsize transf.\!}}
\put(0,0){\textBlack}
\put(90,90){\makebox(0,0){$/\!\!/$}}
\put(93,93){\makebox(0,0)[lb]{\scriptsize$k$}}
\put(80,90){\line(1,0){20}}
\put(90,75){\makebox(0,0){$/\!\!/$}}
\put(93,78){\makebox(0,0)[lb]{\scriptsize$k$}}
\put(80,75){\line(1,0){20}}
\put(0,0){\textBlue}
\put(100,70){\color[rgb]{0.95,0.95,1}\rule{20\unitlength}{40\unitlength}}
\put(100,70){\framebox(20,40){$\mathcal{E}$}}
\put(0,0){\textBlack}
\put(130,90){\makebox(0,0){$/\!\!/$}}
\put(133,93){\makebox(0,0)[lb]{\scriptsize$n$}}
\put(120,90){\line(1,0){20}}
\put(140,90){\line(2,-5){6}}
\put(0,0){\textMagenta}
\put(140,55){\color[rgb]{1,0.95,1}{\rule{20\unitlength}{20\unitlength}}}
\put(140,55){\framebox(20,20){\Magenta{$\mathcal{N}$}}}
\put(0,0){\textBlack}
\put(160,40){\line(-2,5){6}}
\multiput(162,65)(6,0){12}{\line(1,0){3}}
\multiput(138,65)(-6,0){15}{\line(-1,0){3}}
\put(200,67.5){\makebox(0,0)[b]{\scriptsize\Blue{sender}}}
\put(200,62.5){\makebox(0,0)[t]{\scriptsize\ForrestGreen{receiver}}}
\put(170,40){\makebox(0,0){$/\!\!/$}}
\put(173,43){\makebox(0,0)[lb]{\scriptsize$n$}}
\put(160,40){\line(1,0){20}}
\put(140,10){\color[rgb]{1,0.95,0.9}{\rule{20\unitlength}{20\unitlength}}}
\put(140,10){\textcolor{orange}{\framebox(20,20){$\text{\footnotesize id}^{\otimes k}$}}}
\put(160,20){\line(1,0){20}}
\put(170,20){\makebox(0,0){$/\!\!/$}}
\put(173,23){\makebox(0,0)[lb]{\scriptsize$k$}}
\put(180,15){\color[rgb]{0.73,1,0.86}{\rule{20\unitlength}{30\unitlength}}}
\put(180,15){\color[rgb]{0.1,0.37,0.23}{\framebox(20,30){$\mathcal{D}'$}}}
\put(210,32.5){\makebox(0,0){$/\!\!/$}}
\put(205,38){\makebox(0,0)[lb]{\scriptsize$k$ qudits}}
\put(200,32.5){\line(1,0){20}}
\put(222,32.5){\makebox(0,0)[l]{\ForrestGreen{\ForrestGreen{$\ket{\varphi'}$}}}}
\end{picture}}}
\caption{Fully-quantum version of the scheme shown in Fig.~\ref{fig:OurScheme1}.}
\label{fig:OurScheme}
\end{figure}

The decoding operator $\mathcal{D}$ and the correction operators can
be combined as a quantum map $\mathcal{D}'$, see the dashed box in
Fig.~\ref{fig:OurScheme1}.  Furthermore, the sender does not have to
perform a measurement in the generalized Bell basis, but may apply a
unitary transformation that maps the Bell basis to the standard basis,
labeled as ``Bell transf.'' in Fig.~\ref{fig:OurScheme}. Then those
$2k$ qudits can be encoded into $n$ qudits using the quantum map
$\mathcal{E}$.  Hence, Fig.~\ref{fig:OurScheme} shows a fully-quantum
version of our new scheme, showing that our protocol uses the same
type of operations as an EAQECC shown in Fig.~\ref{fig:EAQECC}. We
will also use the same notation $[\![n,k,d;c]\!]_q$ for the
parameters. 

On the other hand, note that we are only transmitting basis states
over the quantum channel, and therefore the protocol is resilient to
arbitrary phase errors.  When following the standard teleportation
protocol, as shown in Fig.~\ref{fig:OurScheme1}, one can replace the
quantum channel by a classical channel.

\begin{figure}[hbt]
\unitlength1.7\unitlength
\centerline{{\begin{picture}(140,130)(-20,-10)
\thicklines
\put(0,0){\vector(1,0){112}}
\put(0,0){\vector(0,1){112}}
\multiput(0,-2)(50,0){3}{\line(0,1){2}}
\put(0,-2){\line(0,1){2}}
\put(0,-3){\makebox(0,0)[t]{\scriptsize$0$}}
\put(20,-2){\line(0,1){2}}
\put(20,-3){\makebox(0,0)[t]{\footnotesize$\frac{1}{5}$}}
\put(25,-2){\line(0,1){2}}
\put(25,-3){\makebox(0,0)[t]{\footnotesize$\frac{1}{4}$}}
\put(50,-3){\makebox(0,0)[t]{\footnotesize$\frac{1}{2}$}}
\put(100,-3){\makebox(0,0)[t]{\scriptsize$1$}}
\put(105,-3){\makebox(0,0)[tl]{\scriptsize$R=k/n$}}
\multiput(-2,0)(0,50){3}{\line(1,0){2}}
\put(-3,0){\makebox(0,0)[r]{\scriptsize$0$}}
\put(-3,50){\makebox(0,0)[r]{\scriptsize$1/2$}}
\put(-2,60){\line(1,0){2}}
\put(-3,60){\makebox(0,0)[r]{\scriptsize$3/5$}}
\put(-2,75){\line(1,0){2}}
\put(-3,75){\makebox(0,0)[r]{\scriptsize$3/4$}}
\put(-3,100){\makebox(0,0)[r]{\scriptsize$1$}}
\put(-3,108){\makebox(0,0)[r]{\scriptsize$\delta=d/n$}}
\thinlines
\put(0,100){\textcolor{magenta}{\line(1,-1){100}}}
\put(12,90){\makebox(0,0)[l]{\textcolor{magenta}{\scriptsize absolute bound \eqref{eq:absoluteBound}}}}
\put(0,50){\textcolor{blue}{\line(2,-1){100}}}
\put(43,30){\makebox(0,0)[l]{\textcolor{blue}{\scriptsize bound \eqref{eq:QuantumSingletonBound} for QECC ($c=0$)}}}
\put(0,75){\textcolor{orange}{\line(4,-3){100}}}
\put(15,64){\makebox(0,0)[l]{\textcolor{orange}{\scriptsize bound \eqref{eq:EASingletonBound} for EAQECC $c=(n-k)/2$}}}
\put(0,50){\textcolor{red}{\line(1,0){50}}}
\put(1,53){\makebox(0,0)[l]{\textcolor{red}{\scriptsize bound \eqref{eq:EASingletonBound1} for EAQECC $c=k$}}}
\put(0,100){\textcolor{forrestgreen}{\line(1,-2){50}}}
\put(13,76){\makebox(0,0)[l]{\textcolor{forrestgreen}{\scriptsize our bound \eqref{eq:OurBound}}}}
\multiput(20,1.25)(0,5){12}{\line(0,1){2.5}}
\multiput(25,1.25)(0,5){10}{\line(0,1){2.5}}
\multiput(50,1.25)(0,5){10}{\line(0,1){2.5}}
\multiput(1.25,60)(5,0){4}{\line(1,0){2.5}}
\end{picture}}}
\caption{Asymptotic bounds (length $n\to\infty$) on the normalized
  minimum distance $\delta=d/n$ as a function of the code rate
  $R=k/n$.}
\label{fig:bounds}
\end{figure}
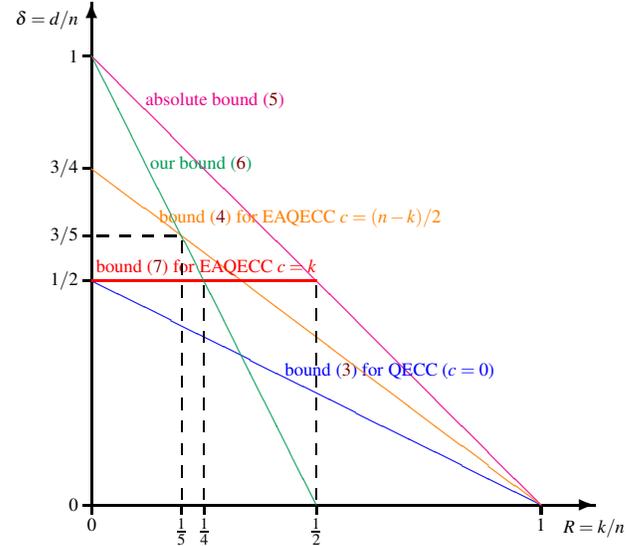

The parameters of our scheme are determined by the classical code $C$.
The Singleton bound for classical codes implies the bound
\begin{equation}\label{eq:OurBound}
d \le n-2k+1
\end{equation}
on the minimum distance of our scheme. It can be achieved whenever the
classical code is an MDS code. In the special case $k=c$, the bound
\eqref{eq:EASingletonBound} implies
\begin{equation}\label{eq:EASingletonBound1}
d \le n/2+1.
\end{equation}
Hence, for $k<n/4$ the bound \eqref{eq:EASingletonBound} is more
restrictive than the bound for our scheme (see also
Fig.~\ref{fig:bounds}).  Even when more maximally entangled states are
used in the original construction of EAQECCs, our scheme has a larger
normalized minimum distance $\delta=d/n$ for a rate $R=k/n$ below a
certain threshold (e.g., $R<1/5$ for $c=(n-k)/2$, see Fig.~\ref{fig:bounds}).

\emph{Examples.}---Classical MDS codes with parameters
$[n,2k,n-2k+1]_q$ are known to exists for $2k\le n\le q+1$.  Using
such a code in our scheme results in parameters
$[\![n,k,n-2k+1;k]\!]_q$, meeting the bound \eqref{eq:OurBound} with
equality.  For prime powers $q\ge 4$, we have in particular classical
MDS codes with parameters $[5,2,4]_q$, yielding a scheme
$[\![5,1,4;1]\!]_q$ using one maximally entangled state. In
comparison, according to \eqref{eq:EASingletonBound}, the minimum
distance $d$ would be at most $3$.  Standard QECC with parameters
$[\![5,1,3]\!]_q$ exist for all $q\ge 2$ and do not require pre-shared
entanglement \cite{Chau97}.

For the case $k=c=1$, i.e., transmitting one qudit with the help of a
single maximally entangled state, we need a classical code with
parameters $[n,2,d]_q$.  For $n\le q+1$, an MDS code $[n,2,n-1]_q$
exists.  Repeating the code $[q+1,2,q]$ $\ell$ times results in a
classical code $[\ell(q+1),2,\ell q]_q$ which is optimal by the
Griesmer bound \cite{Gri60,SoSt65}. The parameters of the resulting
entanglement-assisted communication scheme are $[\![\ell(q+1),1,\ell
    q; 1]\!]_q$, again beating the bound
\eqref{eq:EASingletonBound}. In particular, we get a scheme with
parameters $[\![9,1,6;1]\!]_2$ encoding a single qubit into nine
qubits with the help of one EPR pair \cite{Gra16}.  The normalized
minimum distance $\delta=q/(q+1)$ tends to $1$, while for a fixed
amount $c$ of entanglement, the normalized minimum distance
$\delta=d/n$ is bounded by $1/2$ in \eqref{eq:EASingletonBound}.

\emph{Concluding remarks.}---Quantum codes based on teleportation have
been considered before when studying the entanglement-assisted
capacity of quantum channels \cite[Section III.E]{Bow02}. It was
observed that this results in an entanglement-assisted capacity that
is half the classical capacity of the unassisted quantum channel.  We
are, however, not aware of related results for the finite-length case.

Our scheme beats the quantum Singleton bound
\eqref{eq:EASingletonBound} for quantum communication schemes with a
rate below a certain threshold and uses a smaller amount $c$ of
entanglement than the scheme proposed in \cite{BDH06}.  On the other
hand, when the amount of additional entanglement does not matter,
using $c=n-k$ maximally entangled states in the original scheme
reaches the absolute bound \eqref{eq:absoluteBound}.  It is plausible
to assume that using $c>n-k$ maximally entangled states would not
result in better parameters, as in this case the encoding operation
$\mathcal{E}$ would map $k+c>n$ qudits to a smaller number of qubits.

We conclude by noting that in order to beat the originally stated
quantum Singleton bound for entanglement-assisted quantum-error
correcting codes \eqref{eq:EASingletonBound}, one has to use $c\ge
k$ maximally entangled pairs.  This result, together with further
bounds relating length $n$, dimension $k$, minimum distance $d$, and
the number $c$ of maximally entangled pairs in general
entanglement-assisted schemes can be found in \cite{GHW20}.

\emph{Acknowledgments.}---The author acknowledges fruitful
discussions with Frederic Ezerman, Min-Hsiu Hsieh, Felix Huber,
Ching-Yi Lai, Hui Khoon Ng, Andreas Winter, and Bei Zeng.  The
`International Centre for Theory of Quantum Technologies' project
(contract no. 2018/MAB/5) is carried out within the International
Research Agendas Programme of the Foundation for Polish Science
co-financed by the European Union from the funds of the Smart Growth
Operational Programme, axis IV: Increasing the research potential
(Measure 4.3).

\bibliography{EA_Communication}

\end{document}